\documentstyle[preprint,aps]{revtex}
\input psfig.tex
\begin{document}

\title{Nucleon-Antinucleon Interaction from the Skyrme Model}  

\author{Yang Lu and R.~D.~Amado}

\address{Department of Physics, University of Pennsylvania,
Philadelphia, Pennsylvania 19104}

\date{June 3, 1996}

\maketitle

\begin{abstract}
We calculate the nucleon-antinucleon static potential in the 
Skyrme model using the product ansatz and including some 
finite $N_C$ (number of colors) corrections.  The mid and long
range part of the spin-spin and tensor force are correctly 
given in both iso-spin channels while the central interaction
has insufficient mid-range attraction.  This is a well known
problem of the product ansatz that should be repaired with
better Skyrme dynamics. 
\end{abstract}

\pacs{13.75.Cs, 11.80.Gw}

\section{Introduction}

The Skyrme model \cite{Skyrme} is
an example of what QCD might look like in the classical
or large number of colors ($N_C$) limit \cite{'tHooft,Witten}.
The dynamics of the $SU(2)$ Skyrme model is carried purely by a 
classical pion field.  Hadrons appear as topological solitons in
this non-linear meson field theory.  These are the appropriate degrees
of freedom for the non-perturbative, long wavelength limit of QCD,
and hence for low energy baryon and pion physics. 
The Skyrme model has been applied to the nucleon static 
properties \cite{Adkins} and the nucleon-nucleon interactions 
\cite{Jackson,VinhMau,Ralph,WA} with reasonable success. In the last 
few years, nucleon annihilation has been investigated 
from the Skyrme point of view. Sommermann et al. \cite{SSLK} studied
the dynamics of ungroomed Skyrmion-AntiSkyrmion ($S\overline{S}$) 
collisions. They found that annihilation proceeds quickly
with the creation of a 
coherent pion pulse. This was confirmed by 
Shao, Walet and Amado \cite{Shao}. The notion that annihilation
leads to an intense coherent pion field burst gives reason
for considering annihilation within the classical Skyrme approach.
The idea,  that the annihilation
products, pion and other mesons, come from a coherent wave of meson 
fields arising from soliton-antisoliton dynamics, turns out to be 
very fruitful \cite{batch1,batch2}. Experimental data such as annihilation 
branching ratios among meson types and pion
charge types from low energy annihilation
are well explained with minimal parameters\cite{batch2}. 
Furthermore this picture provides a unified view of annihilation
in which all the channels come from a single process.        

Previous studies of annihilation in the Skyrme context, with the
exception of  \cite{SSLK}, have concentrated on the 
final state mesons.  A full account of the process requires a
description of the initial state 
nucleon-antinucleon interaction and of
the dynamics leading up to annihilation as well. 
In this paper, as a first step in that directions, 
we extend the application of the Skyrme model to 
the interaction of $N\overline{N}$ in the product ansatz. 
(We note that the energy of $S\overline{S}$ in the product ansatz
was studied for two configurations by 
Musakhanov and Musatov \cite{Musakhanov}.)
Phenomenologically, the 
$N\overline{N}$ potential is not as well established as 
the $NN$ potential. At distances less than one fermi, the interaction
is dominated by annihilation. However, at larger distances, 
a meaningful potential can be defined and studied either by
 $G$-transformation on the $NN$ meson exchange potential
 or phenomenologically.  Here we will
compare our Skyrme model 
results to this phenomenology. 
We will see that at large distance, where the product ansatz makes
the best sense, the potentials we find agree qualitatively and,
in most cases, quantitatively with
phenomenological interactions.  At intermediate and short distances,
we do less well, but at these distances    
the product ansatz is not  valid.
However it is still suggestive.
To obtain the interaction at intermediate distances,
we would need to study the full
Skyrme dynamics at these
distances.  For the static $S\overline{S}$ this is 
somewhat more complex 
than in the corresponding $SS$ case studied by Walhout and
Wambach, \cite{W&W}, but is possible and we plan to return to it. 
The full, time dependent,
dynamical $S\overline{S}$ problem is far more difficult
than the $SS$ case 
and is plagued for $S\overline{S}$ 
by numerical instabilities \cite{SSLK,Livermore}. For
all these reasons, and because this paper is a first step, we
begin by exploring the interaction in the product ansatz. 
In order to carry out our study it is
necessary to include $\Delta$ and $\overline{\Delta}$ mixing 
in the potential as was first suggested in the $NN$ case
\cite{old,WA}.

In the context of the product ansatz, we find that
the ungroomed $S\overline{S}$ channel 
studied in \cite{SSLK} is the 
most attractive channel and that it leads to rapid annihilation. 
For non-zero grooming, we find that the $S\overline{S}$ interaction 
can be repulsive. Therefore it seems likely that the
dynamics in groomed channels could   
be very different from that exhibited in 
\cite{SSLK}. Since the physical nucleon 
is represented by an average over 
differently groomed Skyrmions, annihilation in the nucleon-antinucleon
system may 
proceed more slowly than that of ungroomed $S\overline{S}$.

In Sec.~2  we study the interaction energy 
of $S\overline{S}$ as a function of separation and relative
grooming in the product ansatz. We start by studying 
the very simplified case
of Skyrmion and groomed antiSkyrmion on top of each other.  
This is a physically artificial case, but it permits analytic
evaluation and teaches us something, albeit qualitative, about
the dependence of the $S\overline{S}$ interaction on grooming.
We find that for zero separation and zero grooming, the 
$S\overline{S}$ system has zero total energy, 
as we expect.  This is the case
of complete annihilation.  However at relative grooming angle 
of $\pi$, and still zero separation, the product ansatz gives a total 
energy for the $S\overline{S}$ system of four times the
single Skyrmion energy, corresponding to a very repulsive
interaction.  This clearly indicates that the $S\overline{S}$
interaction is a strong function of relative grooming.
Next we study, always in the product ansatz, the interaction energy 
of $S\overline{S}$ at non-zero separation as a function of
grooming. 
We project to the nucleon space by the algebraic methods
of \cite{OBBA} which also include finite $N_C$ corrections.
In Sec.~3 we consider the effects of rotational excitations
by including intermediate states with $\Delta$
and $\overline{\Delta}$. We first evaluate
the corrections to the $N\overline{N}$ potential in 
perturbation theory and then study the effect fully by diagonalizing
in the space spanned by $N$, $\Delta$ and corresponding antiparticles. 
Our results are presented in Sec.~4. 

\section{The interaction energy in the $S\overline{S}$ system 
as a function of separation and relative grooming}

\subsection{The case of an $S$ on top of a groomed $\overline{S}$}

We calculate the energy of the Skyrmion and antiSkyrmion
system using
the Skyrme lagrangian. The density of this lagrangian is given by
\begin{equation}
{\cal L}=\frac{f_\pi^2}{4}{\rm Tr}(\partial_\mu U \partial^\mu U^{+})
+\frac{1}{32e^2}{\rm Tr}(Q_{\mu\nu}Q^{\mu\nu +})+\frac{f_\pi^2}{2}m_\pi^2
{\rm Tr}(U-1),
\end{equation}
where $U$ is a unitary $SU(2)$ valued field and 
\begin{equation}
Q_{\mu\nu}=\left[(\partial_\mu U) U^+, (\partial_\nu U) U^+\right].
\end{equation}
The first term in the lagrangian comes from
the non-linear $\sigma$-model and the second is the Skyrme 
term.  The third term is a  pion mass term and we take 
$m_{\pi}=139$ MeV. We
take the parameters in the lagrangian to have the 
values $f_{\pi} = 93$ MeV and $e=4.76$ \cite{W&W}.  
These values guarantee that the long distance tail of the nucleon-antinucleon
interaction will agree with phenomenology, by virtue of the 
Goldberger-Treiman relation.

We begin by studying the energy in the product ansatz 
for the case of
zero separation. We include the first two terms in the lagrangian.
The mass term is neglected since it does not lead to additional 
understanding of this simple configuration.
Let $U=\exp(i\mbox{\boldmath $\tau$}\cdot\hat{r}F(r))$ 
be the ungroomed Skyrme $SU(2)$ field. The 
ungroomed $\overline{S}$ would be $U^{\dag}$. The rotation 
or grooming matrix $C$ on $\overline{S}$ is 
\begin{equation}
C=\cos(\beta/2)+i\mbox{\boldmath $\tau$}\cdot \hat{n} \sin(\beta/2).
\end{equation}
This corresponds to a grooming rotation through angle
$\beta$ about the $\hat{n}$ axis. 
The product ansatz with this relative grooming is 
\begin{equation}
U_{\rm PA}=UCU^{\dag}C^{\dag}.
\end{equation}
Note that the energy is a function of only the relative grooming and should be
zero with no grooming (since for $\beta=0$, $U_{\rm PA}=1$).

\begin{figure}
\centerline{\hbox{
\psfig{figure=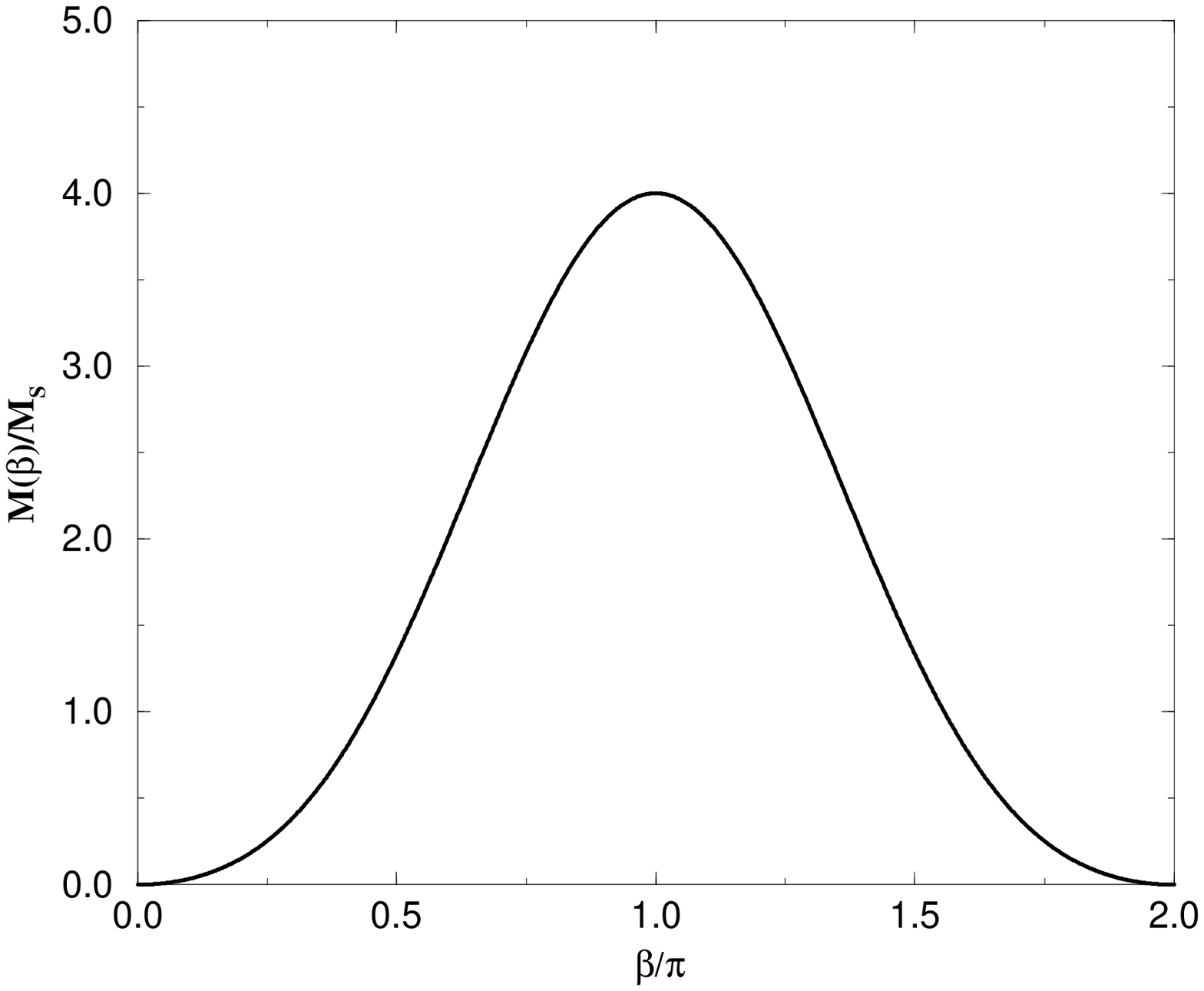,height=3.7in}
}}

\caption{Mass of the $S\overline{S}$ at zero separation and 
with relative grooming angle $\beta$. $M_S$ is
the Skyrmion mass.}\label{Fig.1}
\end{figure}

The energy density is 
\begin{equation}
E=-\frac{1}{4} {\rm tr}{\cal L}_i{\cal L}_i 
-\frac{1}{32} {\rm tr} [{\cal L}_i,{\cal L}_j]^2
\end{equation}
in Skyrme units (energy in $e f_\pi$ and length in $1/(ef_\pi)$).
The chiral (left handed) derivative is 
\begin{equation}
{\cal L}_i= U_{\rm PA}^{\dag}\partial_i U_{\rm PA}.
\end{equation}
Suppose for the Skyrme chiral angle $F(r)$, the 
mass contribution to the $B=1$ skyrmion from the non-linear $\sigma $-model 
term is $M_2$ and from the Skyrme term  $M_4$. 
After some algebra, we arrive at the following result
for the total energy of
the $S\overline{S}$ product ansatz at relative grooming angle $\beta$
 and zero separation.
\begin{equation}
M_{S\overline{S}}(\beta)=\frac{8}{3}\sin^2\left(\frac{\beta}{2}\right)M_2+
\frac{16}{3}\sin^4\left(\frac{\beta}{2}\right)M_4.
\end{equation}
For the profile $F(r)$ which minimizes the $B=1$ Skyrme mass, we
have $M_2=M_4=M/2$ from scaling arguments. Here $M$ is the mass of the 
Skyrmion.
We then have
\begin{equation}
M_{S\overline{S}}(\beta)/M =  \frac{4}{3}\sin^2
\left(\frac{\beta}{2}\right)\left[1+
2\sin^2\left(\frac{\beta}{2}\right)\right],
\end{equation}
which is plotted in Fig.~\ref{Fig.1}.

The maximum occurs at $\beta=\pi$, the maximum grooming, with 
the value $4M$. This 
indicates that the $S\overline{S}$ is quite repulsive at this setting.
Of course for $\beta=0$ or $2 \pi$, the total energy is zero.
This shows that, even for the artificial case of zero separation,
the $S\overline{S}$ energy is very dependent on grooming.

\subsection{$S\overline{S}$ at a separation and a relative grooming}

\begin{figure}
\centerline{\hbox{
\psfig{figure=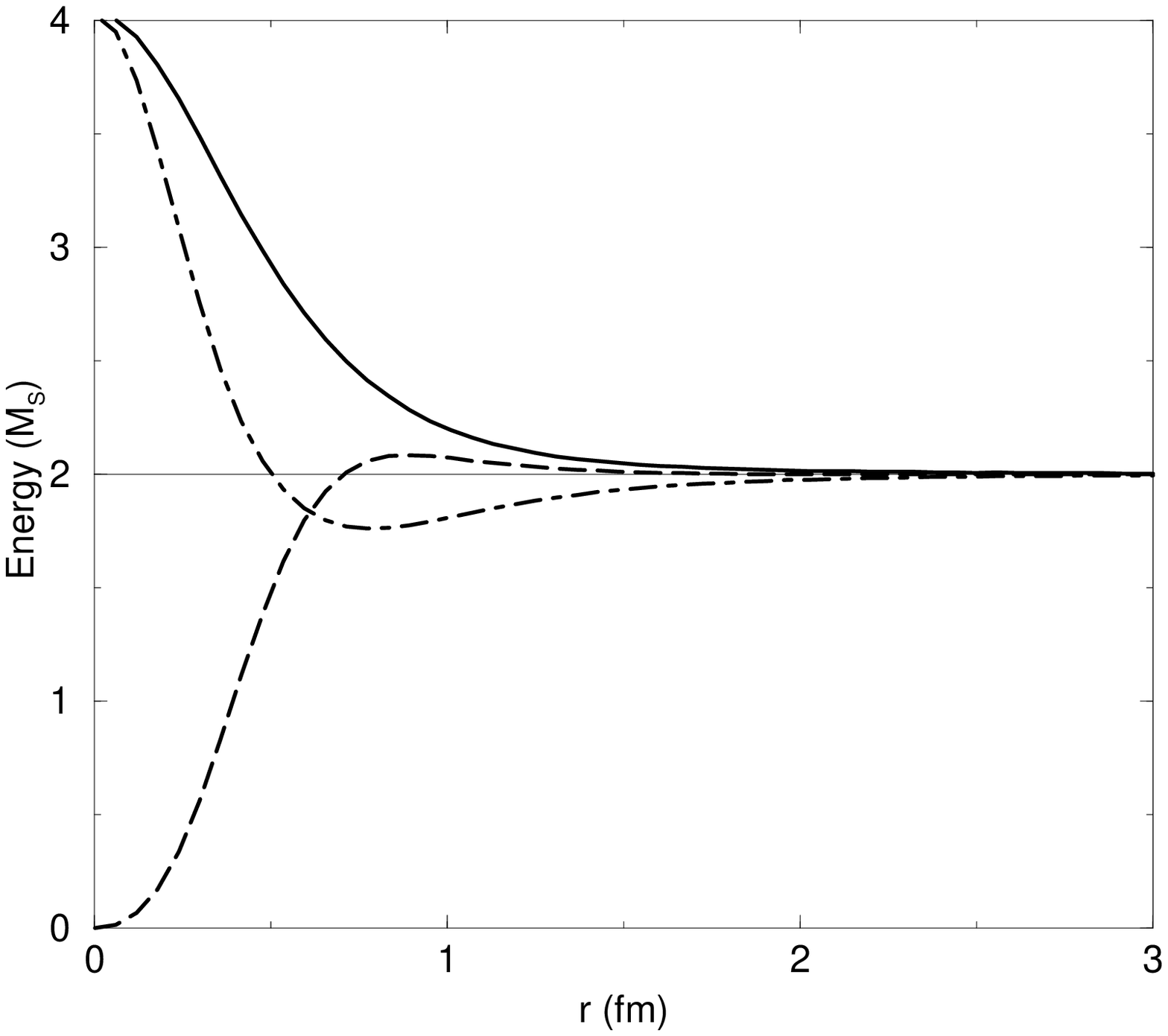,height=3.7in}
}}

\caption{Total energy of the S-antiS system 
as a function of separation for
the configuration HH (dashed line), $x$-$\pi$ (dash-dotted) and
$z$-$\pi$ (solid) in units of the Skyrme mass. 
Note the horizontal line is twice the Skyrmion mass. The maximum 
value of the energy at zero-separation is four times
the Skyrmion mass, as we derived in the analytical result.}\label{Fig.2}
\end{figure}

We now study, in the product ansatz, the energy of the $S\overline{S}$ system 
as a function of the separation and relative grooming. 
This  can only be done numerically. 
We now use the full lagrangian including the finite pion mass
term. We 
put the $S$ and $\overline{S}$ on a 3D lattice with the two solitons
a distance $R$ apart on the $x$ axis. The product ansatz is 
\begin{equation}
U_{\rm PA}(\mbox{\boldmath $r$})=U(\mbox{\boldmath $r$}
-\frac{R\hat{x}}{2})CU^\dagger(\mbox{\boldmath $r$}
+\frac{R\hat{x}}{2}) C^{\dagger},
\end{equation}
where $U(\mbox{\boldmath $r$})$ is the $SU(2)$ field for 
a single Skyrmion and $C$ is the grooming matrix.
In Skyrme units for length ($1/ef_{\pi}=0.45$ fm), the spatial 
extension of lattice we use is 
$20\times 10 \times 10$. We evaluate the derivatives of the 
$U$-field by the two-point difference:
\begin{equation}
\nabla_i U=\frac{1}{2h}\left[ U(\mbox{\boldmath $r$}
+h \mbox{\boldmath $e$}_i)-  
U(\mbox{\boldmath $r$}-h \mbox{\boldmath $e$}_i)\right].
\end{equation}
With $64\times 32 \times 32$ points on the lattice and 
$h=0.001$, we find that, for large separations,
the total energy is within one percent of twice the single
Skyrmion mass. 
We calculate the energy for three interesting configurations:
\begin{enumerate}
\item no grooming ($H\overline{H}$)
\item relative rotation of $\pi$ around $x$-axis ($x$-$\pi$)
\item relative rotation of $\pi$ around $z$-axis ($z$-$\pi$)
\end{enumerate}
The results are shown in Fig.~\ref{Fig.2}.
Recall that the separation is along the $x$ axis.

It is instructive to compare the energy of the $S\overline{S}$
system with the corresponding result for the SS system 
\cite{Jackson,VinhMau,Ralph}.
The $z$-$\pi$ grooming  is the most attractive
configuration for $SS$, while it is the most repulsive for $S\overline{S}$. 
The $x$-$\pi$ grooming, while being the most repulsive for $SS$, is mildly 
attractive for $S\overline{S}$. Finally the ungroomed case is the most 
attractive for $S\overline{S}$
while it is mildly repulsive for $SS$. 
This also leads to speculation about the speed the speed of 
annihilation. In the 
calculation of \cite{SSLK}, the starting configuration is ungroomed and 
annihilation happens very fast.  This may reflect the fact that this
is the 
the most attractive channel. The physical nucleon
is a linear superposition of groomed Skyrmions and 
perhaps in this case 
annihilation will proceed slower than that seen in \cite{SSLK}. 

\subsection{Expansion of the $S\overline{S}$ energy in the 
relative grooming variables}

We now turn to an expansion of the energy in the relative grooming
variables as a first step in obtaining the projection of
the $S\overline{S}$ interaction onto the $N\overline{N}$
interaction.  We  follow the methods developed for obtaining
the $NN$ interaction from the $SS$. 
As in the calculation of  \cite{VinhMau,OBBA} for $SS$, 
the energy for $S\overline{S}$ can be expanded in the variables 
$c_4$ and $\mbox{\boldmath $c$}\cdot\hat{\mbox{\boldmath $R$}}$, with 
the relative grooming matrix 
$C=c_4+i\mbox{\boldmath $\tau$}\cdot\mbox{\boldmath $c$}$
and $\mbox{\boldmath $R$}$ the vector connecting the 
centers of the two solitons.
For the $SS$, the full expansion is 
\begin{equation}
V(\mbox{\boldmath $R$}, C)=V_1+V_2\; c_4^2+V_3 \;
(\mbox{\boldmath $c$}\cdot\hat{\mbox{\boldmath $R$}})^2+V_4\; c_4^4
+V_5\; c_4^2 (\mbox{\boldmath $c$}\cdot\hat{\mbox{\boldmath $R$}})^2+V_6\; 
(\mbox{\boldmath $c$}\cdot\hat{\mbox{\boldmath $R$}})^4
\label{Eq.Vexpand1}
\end{equation}
where $V_i$, $i=1..6$ are functions of $R$.
For $S\overline{S}$, the symmetry of $\mbox{\boldmath $R$}\rightarrow 
-\mbox{\boldmath $R$}$ is
broken by the product ansatz and we need three additional terms 
for a consistent expansion
\begin{equation}
V_{S\overline{S}}=V(\mbox{\boldmath $R$},C)+V_7\; 
c_4 (\mbox{\boldmath $c$}\cdot\hat{\mbox{\boldmath $R$}})+ 
V_8 \; c_4^3 (  \mbox{\boldmath $c$}\cdot\hat{\mbox{\boldmath $R$}})
+V_9\; c_4 (  \mbox{\boldmath $c$}\cdot\hat{\mbox{\boldmath $R$}})^3.
\end{equation}
These terms odd in $\mbox{\boldmath $R$}$ are an 
artifact of the asymmetry of the 
product ansatz and should be discarded.
One can use the symmetrized energy 
$(V_{S\overline{S}}+V_{\overline{S}S})/2$ to extract $V_1$ to
$V_6$, since the $V_7$ to $V_9$ terms drop out in this 
combination.

The six terms in (11) can be expressed in terms of operators in the 
baryon space using the algebraic methods introduced in \cite{OBBA}.
One quantizes each Skyrmion with a $u(4)$ algebra and then the relevant
operators and baryon states are easily constructed in terms of the 
operators of those algebras. The method was developed in \cite{OBBA}
for the $NN$ system, but since each Skyrmion gets its own algebra,
the method can be taken over without alteration to the $N\overline{N}$
system.  The $SS$ or $S\overline{S}$ interaction can be expanded 
in terms of three operators, the identity and the operators $W$ and 
$Z$ given by
\begin{eqnarray}
W&=&T^{\alpha}_{pi}T^{\beta}_{p i}/N_C^2, \nonumber \\
Z&=&T^{\alpha}_{pi}T^{\beta}_{p j}\left[3\hat{R}_i\hat{R}_j-\delta_{ij}\right]
/N_C^2.
\end{eqnarray}
Here $\alpha$ and $\beta$ label 
the two different set of bosons, used to realize
the $u(4)$ algebras.  $T$ is a one-body operator with spin 1 and isospin 1.
The semiclassical (large-$N_C$) limit of these operators can 
be given in terms 
of $\mbox{\boldmath $R$}$ and $C$ as \cite{OBBA}
\begin{eqnarray}
W_{\rm cl}&=& 3 c_4^2-\mbox{\boldmath $c$}^2, \nonumber \\
Z_{\rm cl}&=& 6 {\mbox{\boldmath $c$}}\cdot \hat{\mbox{\boldmath $R$}}
-2\mbox{\boldmath $c$}^2.
\end{eqnarray}

We can therefore 
expand the interaction in $W$ and $Z$, as an alternative to 
Eq.~(\ref{Eq.Vexpand1}). 
\begin{equation}
V(R,C)=v_1(R)+v_2(R)W_{\rm cl}+v_3(R) Z_{\rm cl}+
v_4(R)W^2_{\rm cl}+v_5(R)W_{\rm cl}Z_{\rm cl}+v_6(R)Z^2_{\rm cl}
\end{equation}
The relations between $V_i$ and $v_i$ can be found in 
Eq.~(24) of \cite{OBBA}. The advantage of the algebraic 
method is it allows us to 
study both the large $N_C$ limit, and to include finite $N_C$ effects
explicitly in a systematic way. It also makes taking baryon matrix
elements quite easy.  As in the $SS$ case, we find for
$S\overline{S}$ that the terms quadratic in $Z$ and $W$ are quite small,
and so we neglect them. Hence we can write
\begin{equation}
V=v_1+v_2 W+ v_3 Z.
\label{Eq:potential_wz}
\end{equation}

We can use the algebraic methods of \cite{OBBA} to take the
$N\overline{N}$ matrix element of our interaction.  Keeping
only the leading terms we find
\begin{equation}
V=V_c+V_s(\mbox{\boldmath $\sigma_1$}\cdot
\mbox{\boldmath $\sigma_2$})(\tau_1\cdot\tau_2)
+V_t[3(\mbox{\boldmath $\sigma_1$}\cdot\hat{\mbox{\boldmath $R$}})
(\mbox{\boldmath $\sigma_2$}\cdot\hat{\mbox{\boldmath $R$}})-
\mbox{\boldmath $\sigma_1$}\cdot\mbox{\boldmath $\sigma_2$}]
(\tau_1\cdot\tau_2)
\label{Eq:V_nuconly}
\end{equation}
with 
\begin{equation}
V_c=v_1,\;\;V_s=\frac{v_2 P_N^2}{9},\;\;V_t=\frac{v_3 P_N^2}{9}.\;\;
\end{equation}
Here  $P_N$ is a finite $N_C$ correction factor,
$P_N=1+2/N_C$.
This gives the nucleons only projection of the $S\overline{S}$
interaction.  To obtain the full phenomenological interaction
it is necessary to include the effects of $\Delta$ and 
$\overline{\Delta}$ admixtures that become important as the
baryons approach each other.  We now turn to those admixtures.

\section{Adiabatic Interaction}

The $N\overline{N}$ potential in Eq.~(\ref{Eq:V_nuconly}) is calculated
by projecting Eq.~(\ref{Eq:potential_wz}) to the nucleon degrees of
freedom only.  This is certainly the correct procedure for large
separation.  However as the Skyrmion and antiSkyrmion approach, they can
deform. In terms of the baryon degrees of freedom and $N_C=3$ that means
excitation of the $\Delta$ and $\overline{\Delta}$
intermediate states.  All that is required
to define a $N\overline{N}$ interaction is that the particles be 
 $N\overline{N}$ asymptotically.  They may deform or excite as they
 wish as they interact.  In the $NN$ case we saw that this
intermediate excitation  plays a significant role in the intermediate
range attraction \cite{WA}. 
For neutral atoms, a corresponding virtual excitation 
process leads to the attractive Van der Waals force at large distance. 
For Skyrmions, beside this state mixing,
there is a dynamical distortion that goes beyond
the product ansatz.  This is a crucial part of the $NN$ interaction
\cite{W&W}, but the corresponding  $S\overline{S}$ distortion is beyond
the scope of this paper.  
For the $NN$ system this distortion, coupled with the state mixing,
is crucial for getting the mid-range attraction.  For the nucleon-
antinucleon system we expect similar enhancements of attraction 
coming from the distortion.  The effects of distortion and state mixing
both come in at distances where the product ansatz can be expected
to fail.  Thus our short and mid-range results with the product
ansatz, even including state mixing, should be taken as only
indicative and not as the final word.

As in  the $NN$ system, we first study the effect of mixing $\Delta$ and
$\overline{\Delta}$ on the energy perturbatively and then use the
Born-Oppenheimer method to consider the effect exactly in the limited
subspace.  

\subsection{Perturbation Theory}

We first include the effects of the intermediate states,
$N\overline{\Delta}$, $\Delta \overline{N}$ and $\Delta\overline{\Delta}$
on the nucleon antinucleon 
potential perturbatively.
Since we are using separate $u(4)$ algebras for each Skyrmion,
the results for $NN$ in \cite{WA} can be carried over to 
the $N\overline{N}$ problem, with Eq.~(15) in \cite{WA} being the  
 perturbation correction. 
\begin{eqnarray}
V_{PT}^{(1)}&=& -\frac{Q_N^2}{\delta}\left\{ \left[
\frac{1}{3}Q_N^2P_0^\tau+(\frac{16}{27}P_N^2+\frac{5}{27}Q_N^2)P_1^\tau
\right](v_2^2+2v_3^2)\right.
\nonumber \\
& &+(\mbox{\boldmath $\sigma$}^1\cdot \mbox{\boldmath $\sigma$}^2)\left[
-\frac{1}{18}Q_N^2P_0^\tau+(\frac{16}{81}P_N^2-\frac{5}{162}Q_N^2)P_1^\tau
\right](v_2^2-v_3^2)\\
& & \left. +(3\mbox{\boldmath $\sigma$}^1\cdot\hat{\mbox{\boldmath $R$}}
\mbox{\boldmath $\sigma$}^2\cdot\hat{\mbox{\boldmath $R$}}
-\mbox{\boldmath $\sigma$}^1\cdot\mbox{\boldmath $\sigma$}^2)
\left[-\frac{1}{18}Q_N^2P_0^\tau+(\frac{16}{81}P_N^2
-\frac{5}{162}Q_N^2)P_1^\tau
\right](v_2^2-v_2 v_3) \right\}. \nonumber
\end{eqnarray}
Here $\delta$ is the $N$-$\Delta$ mass 
difference, $P_T^\tau$ is a projection
operator onto isospin $T$, and $Q_N$ is another finite $N_C$ correction
factor with the value $Q_N=\sqrt{(1-1/N_C)(1+5/N_C)}$.
This expresses the leading order correction from state mixing to the 
$N\overline{N}$ interaction of Eq.~(\ref{Eq:V_nuconly}) in 
terms of the $S\overline{S}$
terms of Eq.~(\ref{Eq:potential_wz}).  Recall that unlike the work in \cite{WA}
we are here using the product ansatz to calculate the $S\overline{S}$
interaction rather than a full dynamical scheme.

\subsection{Diagonalization}

We now turn to a full diagonalization of the interaction
in the $N\overline{N}$, $N\overline{\Delta}$, $\Delta\overline{N}$
and $\Delta\overline{\Delta}$ space. This is the Born-Oppenheimer
approximation, and it is valid for $N_C=3$.  There are three energy
scales or time scales in the problem.  The fastest or highest
energy scale comes in rearrangements of the pion field itself.
These we are modeling using the product ansatz and correspond
to energies on the scale of baryon masses.  The intermediate
scale is set by the $N$ $\Delta$ energy difference.  This is an
order $1/N_C$ effect.  Finally the $N\overline{N}$ interaction
is the smallest energy scale and it is determined by the 
matrix diagonalization.

We first need the matrix element of the potential 
Eq.~(\ref{Eq:potential_wz}) in the space of
$N\overline{N}$, $N\overline{\Delta}$, $\Delta\overline{N}$
and $\Delta\overline{\Delta}$ in the angular momentum coupled form. This
has been calculated in
Eq.~(22) of \cite{WA} for the baryon-baryon case and the formula remains 
valid for baryon-antibaryons.
\begin{eqnarray}
\langle I_1I_2LSJT |&v&| I_1'I_2'L'S'JT\rangle \nonumber \\
=&&v_1\delta_{SS'}\delta_{LL'}\delta_{I_1I_1'}\delta_{I_2I_2'}\nonumber\\
&+&\frac{v_2}{9}(-1)^{S+T}\delta_{SS'}\delta_{LL'}
\left\{\begin{array}{ccc}
I_1 & I_2 & S \\
I_2'& I_1'& 1
\end{array}\right\}
\left\{\begin{array}{ccc}
I_1 &I_2 & T \\
I_2'& I_1' &1
\end{array}\right\}
\langle I_1||T^{(11)}||I_1'\rangle
\langle I_2||T^{(11)}||I_2'\rangle \nonumber \\
&+&
\frac{v_3}{9}\sqrt{30}(-1)^{L+L'+S+S'+J+T+I_2+I_1'}
\hat{L}\hat{L'}\hat{S}\hat{S'}
\left(\begin{array}{ccc}
L & 2 & L' \\
0 & 0 & 0
\end{array}\right)
\left\{\begin{array}{ccc}
S & L & J \\
L' &  S' & 2
\end{array}\right\}
\nonumber \\
&\times&
\left\{\begin{array}{ccc}
I_1 & I_2 & S \\
I_2' & I_1' & 1
\end{array}\right\}
\left\{\begin{array}{ccc}
I_1 & I_2 & T \\
I_2' & I_1' & 1
\end{array}\right\}
\left\{\begin{array}{ccc}
I_1 & I_2 & S \\
I_1' & I_2' & S' \\
1 & 1 & 2
\end{array}\right\}
\langle I_1||T^{(11)}||I_1'\rangle
\langle I_2||T^{(11)}||I_2'\rangle
\end{eqnarray}
The relevant reduced matrix elements are
\begin{eqnarray}
\langle N||T^{(11)}||N\rangle &=& -10,\\
\langle \Delta||T^{(11)}||\Delta\rangle &=& -20,\\
\langle N||T^{(11)}||\Delta\rangle &=& -8\sqrt{2}.
\end{eqnarray}

The matrix element of the kinetic part are taken to be very simple,
\begin{equation}
\langle I_1I_2LSJT|K|I_1'I_2'L'S'JT\rangle =
\delta_{I_1I_1'}\delta_{I_2I_2'}\delta_{LL'}\delta_{SS'}
\left(
\delta \left[I_1+I_2-1\right]+\frac{L(L+1)}{2M_{I_1I_2}R^2}\right).
\end{equation}
Here $M_{I_1I_2}$ is the reduced mass, $M_{I_1}M_{I_2}/(M_{I_1}+M_{I_2})$,
with $M_{1/2}=932$ MeV and $M_{3/2}=1232$ MeV. The mass difference is
$\delta M=M_{3/2}-M_{1/2}$. 

For the purpose of comparison, we parametrize the 
full $N\overline{N}$ interaction 
by
\begin{equation} 
V^T_{N\overline{N}}=V_c^T+V_s^T\mbox{\boldmath $\sigma$}^1
\cdot\mbox{\boldmath $\sigma$}^2+
V_t^T\sigma_i^1\sigma_j^2(3\hat{R}_i\hat{R}_j-\delta_{ij}).
\label{Vpara}
\end{equation}
The potentials have explicit isospin dependence due to the
mixing with states of $\Delta$.
To determine the adiabatic potential for $N\overline{N}$, we start at
large $R$ where we have nucleons only. As we move to smaller distance,
we diagonalize the $K+V$ matrix and follow continuously the 
eigenvalue corresponding to the $N\overline{N}$ channel. 
We then subtract the expectation value of $K$ to obtain the adiabatic 
interaction energy as a function of $R$. 

We first consider the case $T=0$. For $J^\pi=0^-$ (note that 
nucleon and antinucleon have opposite intrinsic parity), 
we have three channels
$|N\overline{N}L=0\;S=0\rangle$, $|\Delta\overline{\Delta}L=0\; S=0\rangle$ 
and $|\Delta\overline{\Delta}L=2\; S=2\rangle$. The lowest eigenvalue of
the Hamiltonian $K+V$ should be identified with 
\begin{equation}
\langle L=0\;S=0| V^{T=0}_{N\overline{N}} |L=0\;S=0\rangle 
=
V_c^0-3V_s^0.\label{Eq:V01}
\end{equation}
For $J^\pi=0^+$, there are three channels
$|N\overline{N}L=1\;S=1\rangle$, $|\Delta\overline{\Delta}L=1\; S=1\rangle$ 
and $|\Delta\overline{\Delta}L=3\; S=3\rangle$. The lowest eigenvalue 
should be equated to 
\begin{equation}
\langle L=0\;S=0| V^{T=0}_{N\overline{N}} |L=0\;S=0\rangle 
=
V_c^0+V_s^0-4V_t^0.
\label{Eq:V02}
\end{equation}
We consider one more set of states with $J^\pi=1^-$ and there are six
channels: 
$|N\overline{N}L=0\;S=1\rangle$, 
$|N\overline{N}L=2\;S=1\rangle$,
$|\Delta\overline{\Delta}01\rangle$,
$|\Delta\overline{\Delta}21\rangle$,
$|\Delta\overline{\Delta}23\rangle$ and 
$|\Delta\overline{\Delta}43\rangle$.
The matrix element to identify the lowest eigenvalue with is
\begin{equation}
\langle L=0\;S=1| V^{T=0}_{N\overline{N}} |L=0\;S=1\rangle 
=
V_c^0+V_s^0.\label{Eq:V03}
\end{equation}
From these three linear combinations of $V_c$, $V_s$ and $V_t$
in Eq.~(\ref{Eq:V01}) to Eq.~(\ref{Eq:V03}), 
the potentials in Eq.~(\ref{Vpara}) are 
easily solved for $T=0$.
A similar calculation applies for $T=1$, except that now  
$N\overline{\Delta}$ 
and $\Delta\overline{N}$ channels also appear. 

\section{Result and Discussion}

For each total isospin $T=0,1$, we calculate $V_c^T, V_s^T$ and $V_t^T$
in Eq.~(\ref{Vpara}) as outlined in the last section. 
We then compare these result with the phenomenological potentials 
of Brian-Phillips \cite{BP} and of the Nijmegen \cite{Nijmegen} group.
These are potentials based on meson exchange at large distances
and phenomenology, including an absorptive part to model annihilation,
at small distances. 
The meson exchange part of these potentials for $N\overline{N}$ is obtained
from the corresponding $NN$ potentials by $G$-parity transform --
the contribution of a particular meson for $N\overline{N}$ is 
equal to its part in $V_{NN}$ multiplied by the meson's $G$-parity.
We only compare with the scalar, tensor and spin-spin parts of the
potentials. The spin-orbit force is of higher order in $1/N_C$ 
and we have not calculated it in the Skyrme picture. 
We should note that various cutoffs are used in the Brian-Phillips,
Nijmegen and other similar potentials. As a result at distance 
1 fm or less the strength of the potentials can be significantly
different from their meson exchange value.
In addition, at distance less than 1 fm, the interaction is 
dominated by the absorptive potential of
order 1 GeV. Furthermore, at these short distances the
entire static Skyrme approach, to say nothing of the product
ansatz, is no longer meaningful. Hence we should not place any
faith on comparisons of our results with the phenomenological 
potentials  
around 1 fm or less.  At intermediate distances, between
1 fm and 2 fm, the results from our Skyrme approach to the $NN$
interaction suggest that the product ansatz is 
on the right track, but
that careful comparison with phenomenological potentials requires
a more complete calculation of the Skyrme dynamics.  In particular
we expect the product ansatz to underestimate mid-range attraction,
as it does for $NN$.  This is basically a consequence of the 
variational theorem. 
With these thoughts in mind, let us turn to our results.

Figure 3 shows our results for the $T=0$ part of the central
potential.  We see that the effects of $\Delta$ mixing either in
perturbation theory or full Born-Oppenheimer diagonalization is
significant, but still does not begin to agree with the strong
central, mid-range attraction seen in the phenomenological potentials.
This is the fault of the product ansatz we referred to above.  It
will be important to see if complete Skyrme calculations can
repair this fault. Figure 4 shows the $T=1$ central potential.
Here the effects of $\Delta$ mixing are more striking since
for $T=1$ single $\Delta$ intermediate states are permitted.  
Now we do find some central attraction, but not as much as
is seen phenomenologically.  Note that where they differ, the
full diagonalization result has superior credentials to the
perturbation theory result, and it is the diagonalization result
that is too weak.  Again we must await full Skyrme calculation in
this channel.  Figure 5 shows the $T=0$ spin-spin part of the
interaction.  Except at the smallest distances, the results
are very satisfactory.  Note that this is remarkable, since the
spin-spin interaction is very weak (note the scale in Figure 5),
and hence arises from cancellation of much larger terms.  We
believe it is significant that the Skyrme picture can reproduce
this scale and even the correct sign.  The effects of cancellations
are even more striking in Figure 6 which shows the $T=1$
spin-spin interaction.  Again the order of magnitudes are correct
while the sign depends sensitively on how we do the $\Delta$ 
mixing.  Note that the phenomenological potentials are 
consistent with zero.
Finally in Figures 7 and 8 we show the $T=0$ and $T=1$
tensor potentials.  All calculations of these agree since
they are dominated by one-pion exchange. (Recall that the
Skyrme picture in the product ansatz gets one pion exchange
right.) Hence except for the central attraction, the product
ansatz gives a credible account of the nucleon-antinucleon
potential, and we understand how the product ansatz fails
for the central attraction.  Note that there are no free
parameters on our calculation. 

We have shown that the Skyrme picture with the product ansatz
is a reasonable first step to obtaining the real part of
the nucleon-antinucleon interaction. We also understand
how doing the Skyrme dynamics better can repair the lack
of central attraction we find here.  Hence the next step
is to do that dynamics.  Then combing this picture
of nucleon-antinucleon interactions in the entrance channel
based on the Skyrme model
with our previous work on annihilation channels described
by this model, we hope to have a unified picture of
annihilation based on the large $N_C$, QCD inspired Skyrme
picture.

\section*{Acknowledgement}
This work was supported in part by a grant from the National
Science Foundation.

\begin{figure}
\centerline{\hbox{
\psfig{figure=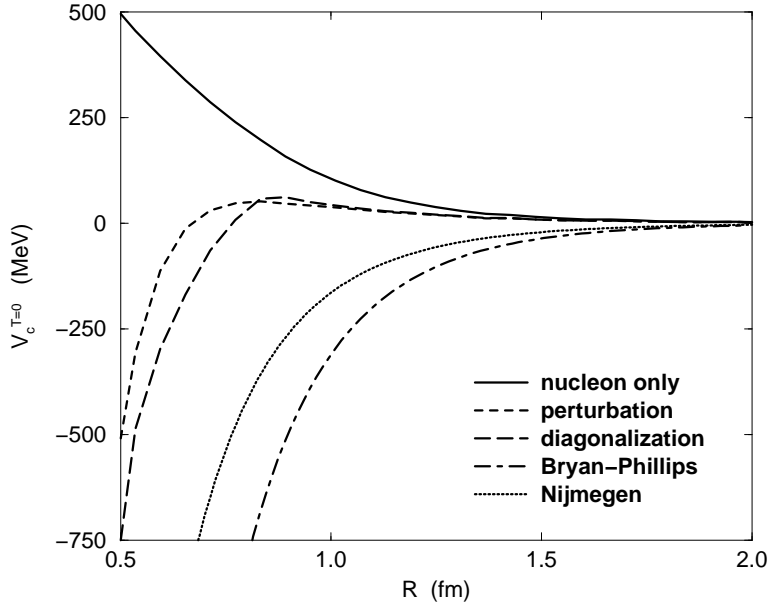,height=3.7in}
}}

\caption{Central potential $V_C^T$ as a function of $R$ in the 
region 0.5-2 fm for  the 
$T=0$ channels. The solid line gives the nucleons only result from 
the product ansatz. The short dashed line is the result of the state
mixing in perturbation theory and the long-dashed line of the 
full Born-Oppenheimer diagonalization.  The meson exchange potentials 
are shown by the dash-dotted line for Bryan-Phillips potential 
{\protect{\cite{BP}}} and by the dotted line for the 
Nijmegen potential {\protect{\cite{Nijmegen}}}.
}\label{Fig.3}
\end{figure}


\begin{figure}
\centerline{\hbox{
\psfig{figure=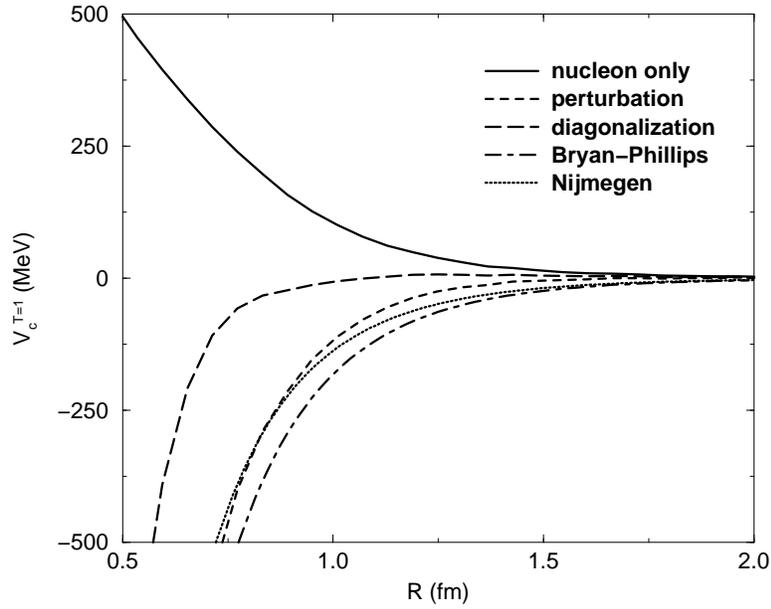,height=3.7in}
}}
\caption{Central potential, same as in Fig.~3 but for $T=1$.}
\label{Fig.4}
\end{figure}


\begin{figure}
\centerline{\hbox{
\psfig{figure=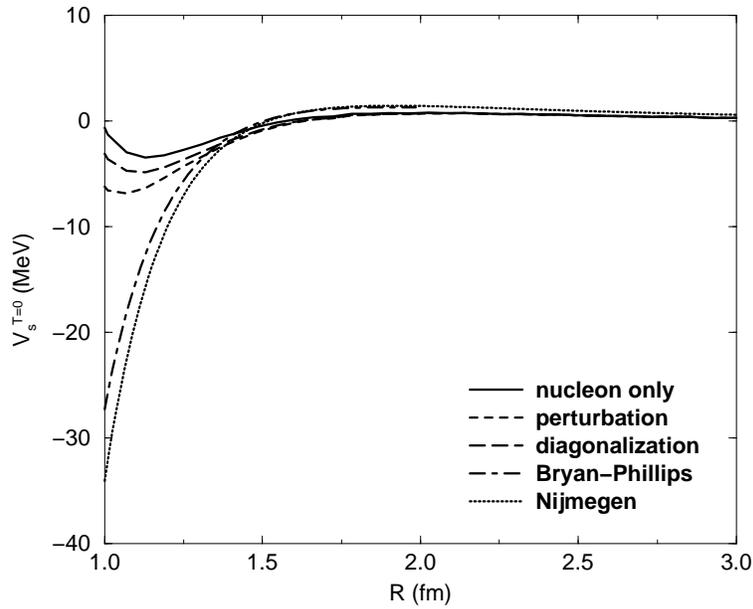,height=3.7in}
}}

\caption{The spin dependent potential $V_s$ as a function of $R$ 
in the region 1-3 fm for 
$T=0$. Labeling of curves is the same as in Fig.~3.}
\label{Fig.5}
\end{figure}


\begin{figure}
\centerline{\hbox{
\psfig{figure=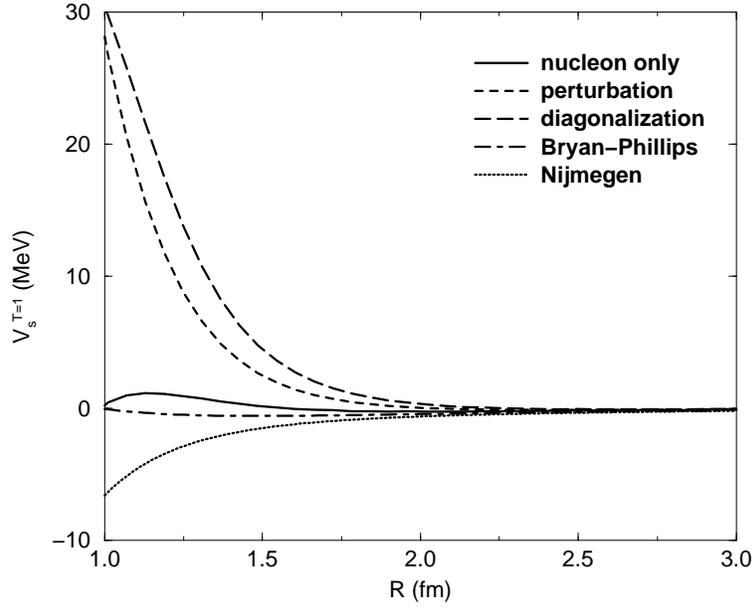,height=3.7in}
}}
\caption{Spin-dependent potential, same as Fig.~5 
but for $T=1$.}
\label{Fig.6}
\end{figure}


\begin{figure}
\centerline{\hbox{
\psfig{figure=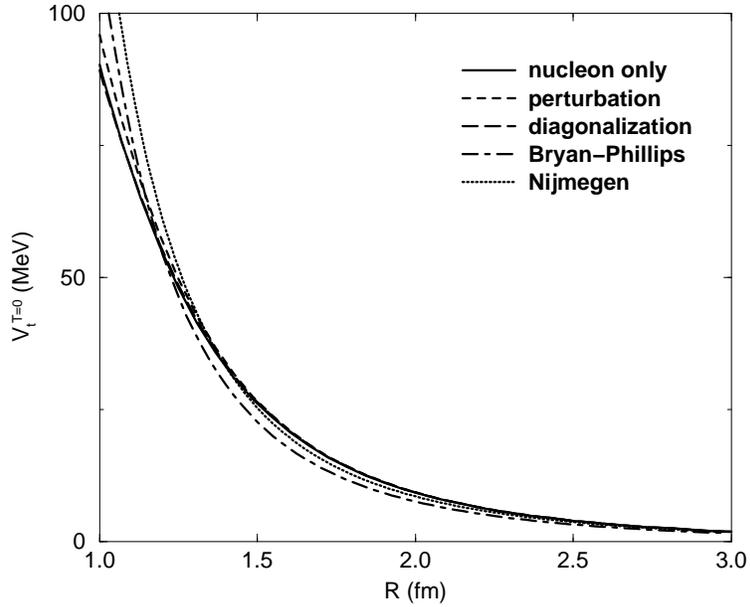,height=3.7in}
}}

\caption{Tensor potential $V_t$ as a function of $R$ 
in the region 1-3 fm for $T=0$. 
Labeling of curves is the same as in Fig.~3.
}\label{Fig.7}
\end{figure}


\begin{figure}
\centerline{\hbox{
\psfig{figure=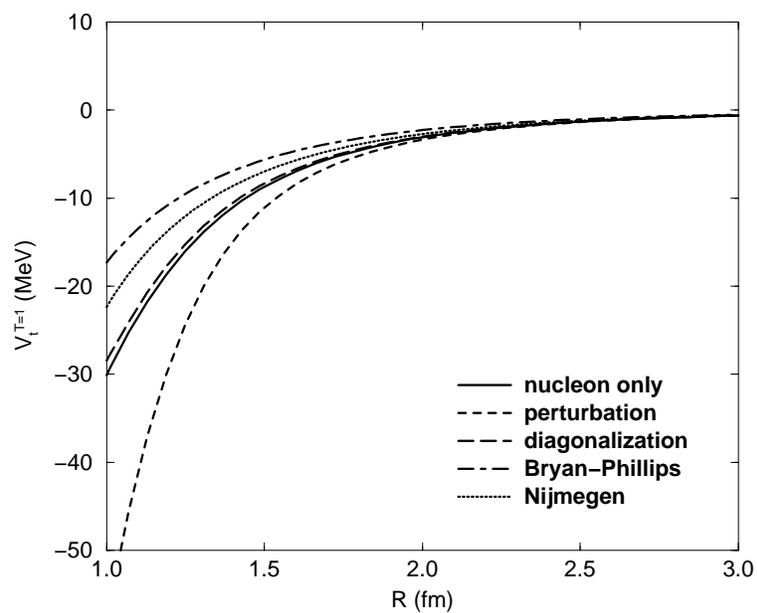,height=3.7in}
}}

\caption{Tensor potential, same as in Fig.~7 but for $T=1$.
 }\label{Fig.8}
\end{figure}

\end{document}